\documentstyle[11pt,newpasp,twoside]{article}
\def\edcomment#1{\iffalse\marginpar{\raggedright\sl#1\/}\else\relax\fi}
\reversemarginpar

\begin{document}
\title{Formation of Young Star Clusters}
\author{Bruce Elmegreen}
\affil{IBM Research Division, T.J. Watson Research
    Center, PO Box 218, Yorktown Hts., NY, 10598, USA,
    bge@watson.ibm.com}

\begin{abstract}
Turbulence, self-gravity, and cooling convert most of the
interstellar medium into cloudy structures that form stars.
Turbulence compresses the gas into clouds directly and it moves
pre-existing clouds around passively when there are multiple
phases of temperature.  Self-gravity also partitions the gas into
clouds, forming giant regular complexes in spiral arms and in
resonance rings and contributing to the scale-free motions
generated by turbulence.  Dense clusters form in the most strongly
self-gravitating cores of these clouds, often triggered by
compression from local stars.  Pre-star formation processes inside
clusters are not well observed, but the high formation rates and
high densities of pre-stellar objects, and their power law mass
functions suggest that turbulence, self-gravity, and energy
dissipation are involved there too.
\end{abstract}

\noindent In Dynamics and Evolution of Dense Stellar Systems, 
IAU Joint Discussion 11, Sydney Australia, July 18, 2004.

\section{Many Scales of Star Formation}

Star formation has many scales. Giant star complexes extend for
$\sim500$ pc along spiral arms and disperse in the interarm
regions. The clouds that form them are usually visible in
galactic-scale HI surveys, and their cores are visible in CO
surveys (Grabelsky et al. 1987). Many of these clouds are mildly
self-bound by gravity (Elmegreen \& Elmegreen 1987; Rand 1993), so
they are like any other star-forming clouds: virialized,
supersonically turbulent, and capable of producing stars in
perhaps several generations with an overall efficiency of
$\sim10$\%.  The star formation process itself is confined to the
densest cores of these clouds, where gravity is strong and thermal
pressure is weak. Between these extremes of scale, the gas
temperature decreases and the molecular content increases, but the
physical processes that cause stars to form in aggregates do not
appear to change much.  These processes are a combination of
multi-scale and repetitive compressions from supersonic turbulence
and self-gravity, energy dissipation through shocks and magnetic
diffusion, and contraction or collapse from overwhelming
gravitational forces.  Some of the complexity of star-formation
dynamics is shown in the simulations by Bate, Bonnell, \& Bromm
(2003).

\section{Scale-dependent Morphologies}

Corresponding to the many scales of star formation,
self-gravitating clouds have a wide range of masses, from
$\sim10^7$ M$_\odot$ to less than 1 M$_\odot$ in our Galaxy. What
a cloud produces is called a star cluster only if its mass exceeds
$\sim100$ M$_\odot$ (Lada \& Lada 2003). Other than this, there is
no characteristic or dominant mass for clouds or clusters, only
power law distributions, so most star-forming regions are similar
except for size. Size determines velocity dispersion and density
for a common background pressure, and density variations lead to
important morphological differences through two dimensionless
ratios: the dynamical time divided by the evolution time of stars,
and the dynamical time divided by the shear time in the local
galaxy. The largest clouds take a short time, in relative terms,
to form most of their stars: just 1 or 2 dynamical times like
nearly every other cloud. But these largest clouds take a long
time, in absolute terms, to do this, $\sim40$ My in the case of
Gould's Belt, and by then the oldest populations have lost their
most massive members to stellar evolution, making the complexes
look relatively dull (Efremov 1995). The largest clouds are also
the most severely affected by shear, making them look like
flocculent spiral arms or spiral arm spurs (Kim \& Ostriker 2002).
These morphological differences disguise the fact that the
physical processes of star and cluster formation are very similar
on all scales.

Galactic-scale stellar dynamical processes can lead to the
collection of gas into spiral density waves and resonance rings.
Then the largest clouds are somewhat uniformly distributed along
the length of the stellar structure with a characteristic
separation equal to $\sim3$ times the arm or ring thickness. What
happens here is that clouds form by asymmetric gravitational
instabilities with a converging flow along the length of the
structure. Typically shear and galactic tidal forces are low in
these regions, allowing the clouds to form in gas that would
otherwise be stable (Rand 1993; Elmegreen 1994).

\section{Power Spectra}

When there are no galactic-scale structures, the gas appears
completely scale-free, as in the Large and Small Magellanic Clouds
(Stanimirovic et al. 1999; Elmegreen, Kim, \& Staveley-Smith
2001). Power spectra of the emission or absorption from this gas
have power laws with a slope similar to that for velocity power
spectra in incompressible turbulence, namely $\sim-2.8$ in
two-dimensions (St\"utzki et al. 1998; Dickey et al. 2001).
Incompressible turbulence has the Kolmogorov spectrum with a slope
of $-8/3$. Why the {\it column density} structure in a medium that
is supersonically turbulent should have about the same power
spectrum as the {\it velocity} structure in incompressible
turbulence is somewhat of a mystery, unless it is partly
coincidence.  The power spectrum of turbulent velocities varies by
only a small amount, from $-8/3$ to $-3$ (in 2D), as the motion
varies from incompressible to shock-dominated. Thus even the most
extreme cloud formation scenarios, where all clouds are shock
fronts, would have a power spectrum similar to incompressible
turbulence. In addition, some of the gas structure could result
from entrainment of many tiny clouds in the larger-scale turbulent
velocity field (Goldman 2000). Entrainment means density is a
passive scalar, and then density power spectra are the same as
velocity power spectra. Third, expanding shells make dense gas,
and these introduce a $-3$ component to the power spectrum because
of their sharp edges.  The result is a mixture of processes and
innate power spectra.  This is why widely diverse morphologies
ranging from flocculent dust spirals in galactic nuclei
(Elmegreen, Elmegreen, \& Eberwein 2002) to shells and holes in
the LMC or SMC (Kim et al. 1999; Stanimirovic et al. 1999;
Elmegreen et al. 2001; Lazarian, Pogosyan, \& Esquivel 2002) all
have about the same overall power spectrum.

\section{Stars Follow the Gas}

Star formation structures, such as clusters and flocculent spiral
arms, have hierarchical geometries (Feitzinger \& Galinski 1987;
Gomez et al. 1993; Elmegreen \& Elmegreen 2001; Zhang, Fall, \&
Whitmore 2001) and power-law power spectra (Elmegreen, Elmegreen,
\& Leitner 2003; Elmegreen et al. 2003) that are nearly identical
to those of the gas. Star formation also has a duration that
scales with the region size in the same way as the turbulent
crossing time scales with size (Efremov \& Elmegreen 1998). These
similarities between star formation and turbulent gas imply that
star formation follows the gas to first order, i.e., that
turbulence controls the star formation density, rate, and
morphology.

This control apparently extends to small scales too, perhaps down
to individual binary stars (Larson 1995), as the protostars in
clusters sometimes have their own hierarchical structure (Motte,
Andr\'e, \& Neri 1998; Testi et al. 2000).  The large formation
rates and high densities of embedded protostars also suggest that
turbulence compresses the gas in which they form (Elmegreen \&
Shadmehri 2003).

\section{Triggering}

Closer examination also shows a second-order effect: that a fairly
high fraction of star formation is also triggered inside
pre-existing clouds by external pressures unrelated to the clouds
and to the pressures of the current generation.  These processes
are revealed by the wind-swept appearance of many cluster-forming
clouds (e.g., de Geus 1992; Bally et al. 1987) and by the
proximity of cluster-forming cores to external HII regions
(Yamaguchi et al. 1999; Walborn et al. 1999; Heydari-Malayeri et
al. 2001; Yamaguchi et al. 2001a,b; Deharveng et al. 2003). What
is probably happening is that supersonic turbulence and
entrainment in a multi-phase ISM produce the basic cloudy
structure, and then unrelated pressure fluctuations in the
environment trigger star formation in these structures (Elmegreen
2002). Presumably there would still be star formation without the
triggers, but with an average rate per cloud that is less because
of the lower cloud densities, and a number density of active
clouds that is greater because of the more dispersed nature of the
dense sub-regions. The influence of pressurized triggering on the
overall star formation rate in a galaxy is not known, but the
universal scaling of star formation rate with average density or
column density (Kennicutt 1998) suggests that any direct influence
is weak.  Star formation is probably saturated to the maximum rate
allowed by the density structure in a compressibly turbulent
medium (Elmegreen 2002).

\section{Size of Sample Effects}

The stochastic nature of turbulence is also reflected in the
formation of star clusters, which show a random size-of-sample
effect with regard to maximum mass.  This appears in several ways:
the most massive stars in a cluster increase with the cluster mass
(Elmegreen 1983), the most massive clusters in a galaxy increase
with the number of clusters (Whitmore 2003; Billett et al. 2002;
Larsen 2002), and the most massive clusters in a logarithmic age
interval increase with the age (Hunter et al. 2003).  In all
cases, the slopes of these increases are determined exclusively by
the mass function through the size of sample effect: bigger
regions sample further out in the tail of the distribution and
have more massive most-massive members. There is apparently no
physical effect or physical parameter that has yet been found to
determine the most massive member of a population. This is true
even for individual stars (Massey \& Hunter 1998; Selman et al.
1999) although stellar radiation pressure and winds could limit
the stellar mass once it gets large enough (Yorke \& Sonnhalter
2002; but see McKee \& Tan 2003).

Similarly, the ISM pressure should limit the cluster mass,
considering that a cluster is recognized only if its density
exceeds a certain value (depending on the sensitivity of the
observation), and the density, mass and pressure are related by
the virial theorem with a boundary condition. Nevertheless, this
pressure limit for massive clusters has not been seen yet. It
would appear as a drop-off at the upper end of the cluster mass
function in a very large galaxy (sampling lots of clusters) with a
low pressure (such as a giant low-surface brightness galaxy). Most
galaxies have their sample-limiting mass comparable to or less
than their pressure-limiting mass. Dwarf starburst galaxies are an
extreme example of this because they have very few clusters
overall and yet some high pressure regions. Dwarf galaxies do
indeed have an erratic presence of massive clusters, some of which
may be related to galaxy interactions (Billett et al. 2002).

\section{Summary}

Most stars form in clusters (Carpenter 2000; Lada \& Lada 2003)
and many of these clusters are close enough to high-pressure
regions to look triggered.  Triggering seems necessary because the
dynamical pressures inside clusters are several orders of
magnitude larger than the ambient interstellar pressure.  The high
pressure state of a cluster is an obvious remnant of its birth,
but clues to the origin of the pressure are lost once the gas
disperses and the stellar orbits mix.  {\it The primary
distinction between the formation of standard ``open clusters''
and the mere aggregation of stars in a compressibly turbulent
medium is probably this last step of triggering}.  HII regions did
not compress gas to make Gould's Belt, but they did compress gas
to make the Trapezium cluster in Orion.

The masses and positions of the clouds that are compressed into
clusters seem to be the result of interstellar turbulence and
shell formation.  Turbulence structures the gas in two ways: by
directly compressing parts of it through random large-scale flows,
and by moving pre-existing clouds around passively. This duality
of processes follows from the multi-phase nature of the ISM and
from the presence of self-gravity. Combine these with pervasive
pressure bursts from massive stars and the result is a mode of
star formation dominated by dense clusters.

\end{document}